\documentstyle[preprint,aps,tighten]{revtex}
\begin{document}
\draft
\title{Screened thermonuclear reactions and predictive stellar evolution of
detached double-lined eclipsing binaries}
\author{Theodore E. Liolios$^{1}$ \thanks{%
Corresponding author: www.liolios.info}, Theocharis S. Kosmas$^{2}$}
\address{$^{1}$Hellenic Military University, Hellenic Army Academy\\
Department of Physical Sciences, Vari 16673, Attica, GREECE\\
$^{2}$Theoretical Physics Division, University of Ioannina, Ioannina 45110,
GREECE}
\maketitle

\begin{abstract}
The low energy fusion cross sections of charged-particle nuclear reactions
(and the respective reaction rates) in stellar plasmas are enhanced due to
plasma screening effects. We study the impact of those effects on predictive
stellar evolution simulations for detached double-lined eclipsing binaries.
We follow the evolution of binary systems (pre-main sequence or main
sequence stars) with precisely determined radii and masses from 1.1M$_{\odot
}$ to 23M$_{\odot }$ (from their birth until their present state). The
results indicate that all the discrepancies between the screened and
unscreened models (in terms of luminosity, stellar radius, and effective
temperature) are within the observational uncertainties. Moreover, no
nucleosynthetic or compositional variation was found due to screening
corrections. Therefore all thermonuclear screening effects on the
charged-particle nuclear reactions that occur in the binary stars considered
in this work (from their birth until their present state) can be totally
disregarded. In other words, all relevant charged-particle nuclear reactions
can be safely assumed to take place in a vacuum, thus simplifying and
accelerating the simulation processes.
\end{abstract}


The most effective tests of stellar evolution theory are provided by the sun
and binary stars because their masses, radii, effective temperatures and
luminosities can be determined with high accuracy\cite{Schroeder97,Pols97}.

Recently several papers appeared\cite{Young1,Young2,Young3} which applied
the {\it TYCHO}\cite{tycho60}\ stellar evolution code in an effort to assess
its predictive capability by the critical evaluation of its assumptions. Ref.%
\cite{Young1} presented baseline results from stellar models calculated
using the {\it TYCHO} code, which were then tested against a subset of
double-lined eclipsing binaries\cite{Andersen}. Other papers followed by the
same group which elaborated on hydrodynamic regions\cite{Young2} and/or
included more realistic physics\cite{Young3}. According to those studies the
stellar evolutionary history of the sun (as well as of other stars) as
predicted by solar evolution simulations is not unique. The main results are:

1) A model which fits the present day sun may be substantially inaccurate
for other evolutionary stages and stars. The sources of this uncertainty are
mainly the mixing length theory and the uncertainties in the solar
abundances.

2) Diffusion has insufficient time to work on the pre-main sequence path
(pre-MS) and does not affect the tracks

3) Hydrodynamic mixing (inertial-driven mixing) plays an important role in
solar evolution only in the pre-MS path where there exists a small
convective core (during partial CN burning).\newline
Although Refs.\cite{Young1,Young2,Young3} have studied in detail the effects
of mixing and settling/diffusion on stellar predictive simulations, they
have not investigated another source of uncertainty, which results from the
fact that nuclear reactions in stellar interiors are influenced by the
plasma which surrounds the reacting nuclei. Actually, neither version of the 
{\it TYCHO} code (6.0, 6.11) which are publicly available is equipped with
screening corrections for the reaction rates, while the most recent
application of {\it TYCHO}\cite{Young3} despite its many improvements
doesn't discuss this effect, either. It is obvious that thermonuclear
screening might be the source of uncertainty to all efforts to simulate and
predict the current state of double-lined eclipsing binaries, and that is
why the present work is timely and relevant.

It should be emphasized that the effects of screening corrections on
helioseismological aspects of stellar evolution is another major issue which
has been studied extensively in Refs.\cite
{Fiorentini2000,Tsytovich2000,Weiss2001} (and references therein). According
to these papers the existence of screening effects in the solar interior can
be proved by helioseismology\cite{Fiorentini2000} since the absence of such
relevant corrections in computer simulations yields solar sound speed
profiles which are incompatible with the seismic ones established by
helioseismology (a discrepancy of the order of $\sim 1\%$\cite
{Fiorentini2000}). 

Moreover, Ref.\cite{Weiss2001} proved that good agreement with the seismic
sound profile is obtained from screening factors which do not deviate by
more than $\pm 10\%$ from the standard screening prescription (i.e.
Salpeter's formalism \cite{Salpeter}). Therefore, since the screening model
TYCHO\ has been recently equipped with (i.e. Mitler's one, see below)
practically coincides\cite{Fiorentini2000} with Salpeter's one for solar
conditions, all TYCHO solar evolution simulations (e.g. Ref. \cite{Young3})
can now be more compatible with helioseismological and solar neutrino data.

The enhancing influence of stellar plasmas on astrophysical thermonuclear
reactions (and on the respective thermonuclear reaction rates) has been
studied by many authors (see for example Refs.\cite
{Salpeter,mitler,dzitko,shav,darshaviv,ricci,lioliosprc2000,Bahcall98,lioliosprc2001}%
, and references therein) who derive corrective factors (known as Screening
Enhancement Factors:SEFs) by which the reaction rates are multiplied in
order to take into account screening effects.

We can define the SEF by using the usual definition\cite{Claytonbook} of the
binary thermonuclear reaction rate $r_{ij}\,$per particle pair\thinspace $%
\left( i,j\right) $

\begin{equation}
r_{ij}\sim \int_{0}^{\infty }S\left( E\right) P\left( E\right) \exp \left( -%
\frac{E}{kT}\right) dE
\end{equation}
where the penetration factor $P\left( E\right) $ multiplied by the
astrophysical factor $S\left( E\right) $ in the $s$-wave cross section
formula

\begin{equation}
\sigma \left( E\right) =\frac{S\left( E\right) }{E}P\left( E\right)
\end{equation}
is given by the WKB method:

\begin{equation}
P\left( E\right) =\exp \left[ -\frac{2\sqrt{2\mu }}{\hbar }%
\int_{R}^{r_{c}\left( E\right) }\sqrt{V\left( r\right) -E}dr\right]
\label{pwkb}
\end{equation}
and the classical turning point $\left( r_{c}\right) \,$is the distance
between the colliding nuclei at which the potential energy of the
interaction $V_{SC}\left( r\right) \,$equals their kinetic energy $E$ so
that:

\begin{equation}
V_{sc}\left( r_{c}\right) =E
\end{equation}
The screening enhancement factor $\left( f_{nr}\right) $ for non-resonant
charged-particle reactions will be given by the ratio of the screened
penetration factor $P_{SC}\left( E\right) \,\,$with respect to the
unscreened one $P_{NSC}\left( E\right) \,\,\,$evaluated at the most
effective energy of interaction\thinspace $E_{0}$ so that:

\begin{equation}
f_{nr}\left( E_{0}\right) =P_{SC}\left( E_{0}\right) /P_{NSC}\left(
E_{0}\right)  \label{ppnr}
\end{equation}
(For a detailed analysis of deriving SEFs see Ref.\cite{lioliosprc2000})

There is a variety of models, each of which has inherent limitations, while
some of them have been the subject of intense controversy\cite{bahcall5}.
Some very widely used screening models are Salpeter's weak screening (WS)
model (S)\cite{Salpeter}, Graboske-DeWitt's\cite{Graboske73} and Mitler's
one (M)\cite{mitler}. Actually these models are frequently used in solar
evolution codes giving quantitative estimates of the neutrino flux
uncertainties associated with the screening effect(see for example Refs.\cite
{Salpeter,mitler,dzitko,shav,darshaviv,ricci,lioliosprc2000,Bahcall98,lioliosprc2001}
and references therein). Others studies\cite{Ichimaru} have dealt with
strongly degenerate regimes where screening effects are extremely important.
However, the effects of thermonuclear screening on the evolution of detached
eclipsing binaries (focusing on the reaction rates and the relevant
nucleosynthesis) has never been studied in detail.

{\it TYCHO} has been recently equipped by one of the authors (T.L.) with
Mitler's\cite{mitler} screening corrections for nuclear reaction rates which
offers the opportunity to exhaustively investigate: a) the relevance of
thermonuclear screening in stellar evolution and nucleosynthesis, and b) the
error committed by not including such correction in relevant simulations\cite
{Young1,Young2,Young3}.

The {\it TYCHO} code has been analyzed in detail in Refs\cite
{Young1,tycholiolios} therefore we only briefly present the parameters of
the version that we used. Actually, in all models created in the present
paper we used the parameters adopted in Ref.\cite{Young1} for consistency
which will facilitate the study of the effects of thermonuclear screening.
Namely, we used Timmes \& Arnett's equation of state\cite{Young1,Timmes},
opacities from Iglesias \& Rogers (1996)\cite{Iglesias96} and Kurucz (1991)%
\cite{Kurucz91} for a solar abundance pattern from Anderse-Grevesse 1989\cite
{AG89}. Although there are more recent solar abundances\cite
{Grevesse98,Lodders} which could also be used to produce compatible opacity
tables we have decided to use the same mix\cite{AG89} as that of Ref.\cite
{Young1,Arnettbook} both in the construction of our initial models and the
derivation of the OPAL Rosseland mean opacity tables. Following Ref.\cite
{Young1} we have selected the lower mass limit of our stars to be well above
the limit of validity for our equation of state\cite{Young1} $\left(
1.1M_{\odot }<M\right) .$

In our simulations we apply the standard Schwartzchild convective theory\cite
{Claytonbook} assuming that a mixing length $(l)$ exists which is
proportional to the pressure scale height $(H_{p})$ so that: $a_{p}=l/H_{p}$%
. Ref.\cite{Young1} assumed that $a_{p}=1.6,$while comparisons\cite{Maeder}
between models with mixing lengths of $a_{p}=1.5$ and $a_{p}=1.9$ clearly
support a value of $a_{p}=1.9\pm 0.1\,$. We have experimented using various
mixing lengths in our simulations including of course the value of $%
a_{p}=1.6,$ so that we are compatible with Ref.\cite{Young1}. No correlation
was found between the mixing length and the screening enhancement factor in
our models.

In this work we are interested in isolating the effects of thermonuclear
screening, therefore we follow the methodology used in Ref.\cite{Young1}
avoiding optimization of all relevant parameters. Thus, we turn convective
overshooting off while we also disregard all mass loss effects (the effects
of the latter being minor anyway\cite{Young1}). Rotational mixing was also
turned off while the inertial-driven mixing (TYCHO v.6.11\cite{Young2}) was
not taken into account.

In Ref.\cite{Young1,Young3} a sample of stars ($1.1M_{\odot }<M<23M_{\odot }$%
) was used from Andersen's\cite{Andersen} double-lined eclipsing binaries
which had precisely determined masses and radii. We used the same sample so
that a direct comparison of the same code with and without thermonuclear
screening corrections can be made.

The method followed in this work is plausible and simple: We follow the
individual evolution of each member of the binary systems studied in Refs.%
\cite{Young1,Young3} from their birth until their present state (using the
ages indicated in Refs.\cite{Young1,Young3}) with and without screening
corrections for the nuclear reaction rates and we observe the differences
between the two parallel tracks. Note that those ages have been obtained by
following a plausible fitting procedure which fits the derived models to the
observational data without using optimization for masses or abundances.
Other methods (e.g. Ref.\cite{Ribas}) have adopted helium and heavy element
variation but such a complex procedure would not serve any purpose in our
study which aims at isolating the effect of screening rather than deriving
the most accurate stellar evolution models.

The results indicate that all the discrepancies between the screened and
unscreened models (in terms of luminosity, stellar radius and effective
temperature) are within the current observational uncertainties. Moreover,
no nucleosynthetic or compositional variation was found due to screening
corrections. Therefore all thermonuclear screening effects on the
charged-particle nuclear reactions that occur in the binary stars considered
in this work (from their birth until their present state) can be totally
disregarded. In other words, all relevant charged-particle nuclear reactions
can be safely assumed to take place in a vacuum, thus simplifying and
accelerating the simulation processes.

{\bf Acknowledgments:} This work was financially supported by {\it the
Hellenic General Secretariat for Research and Technology (www.gsrt.gr)}
under program {\it PENED 2003} {\it (Space Sciences \& Technologies)}.


\begin{references}
\bibitem{Schroeder97}  K.P.Schroeder, O.R.Pols, P.P.Eggleton,
Mon.Not.R.Astron.Soc. 285$\left( 1997\right) 696$

\bibitem{Pols97}  O.R.Pols, C.A.Tout,K.P.Schroeder, P.P. Eggleton,
J.Manners, Mon.Not.R.Astron.Soc. {\bf 289}$\left( 1997\right) 869$

\bibitem{Young1}  P.A.Young, E.E.Mamajek, D.Arnett, J.Liebert, ApJ ${\bf 556}%
\left( 2001\right) 230$

\bibitem{Young2}  P.A.Young, K.A.Knierman, J.R.Rigby, D.Arnett, ApJ ${\bf 595%
}\left( 2003\right) 1114$

\bibitem{Young3}  P.A.Young, D.Arnett,Ap.J. ${\bf 618}(2005)908$

\bibitem{tycho60}  The website of {\it TYCHO }maintained by the university
of Arizona is: http://chandra.as.arizona.edu/\symbol{126}%
dave/tycho-manual.html

\bibitem{Andersen}  J.Andersen, A\&A Rev., ${\bf 3}\left( 1991\right) 91$

\bibitem{Fiorentini2000}   G. Fiorentini, B. Ricci, F.L. Villante,
Nucl.Phys.Proc.Suppl. {\bf 95 }(2001) 116-122

\bibitem{Tsytovich2000}  V.N.Tsytovich, A\&A Letters, 356$\left( 2000\right)
L57$

\bibitem{Weiss2001}  A.Weiss A, M.Flaskamp, V. Tsytovich A\&A, 371$\left(
2001\right) $1123

\bibitem{Salpeter}  {E. E. Salpeter, Aust. J. Phys. {\bf 7}, 373 (1954).}

\bibitem{mitler}  {H. E. Mitler, Astrophys. J. {\bf 212}, 513 (1977).}

\bibitem{dzitko}  {H. Dzitko {\it et al.}, Astrophys. J. {\bf 447}, 428 }%
(1995).

\bibitem{shav}  {N. Shaviv and G. Shaviv, Astrophys. J. {\bf \ 468}, 433
(1996).}

\bibitem{darshaviv}  {A. Dar and G. Shaviv, Astrophys. J. {\bf 468}, 933
(1996).}

\bibitem{ricci}  B.Ricci, S.Degl'Innocenti, G.Fiorentini, Phys.Rev.C. {\bf 52%
}, 1095$\left( 1995\right) $

\bibitem{lioliosprc2000}  T.E.Liolios, Phys.Rev.C {\bf 61}, 55802$\left(
2000\right) $

\bibitem{Bahcall98}  J.Bahcall, X.Chen, M.Kamionkowski, Phys.Rev.C. $%
57\left( 1998\right) 2756$

\bibitem{lioliosprc2001}  T.E.Liolios, Phys.Rev.C {\bf 64}, 018801$\left(
2001\right) $

\bibitem{Claytonbook}  D.D.Clayton, Principles of Stellar Evolution and
Nucleosynthesis, McGraw-Hill Book Company $\left( 1968\right) $

\bibitem{bahcall5}  J.N.Bahcall, L.Brown, A.Gruzinov,R.Sawer,
Astron.Astrophys. {\bf 383}, 291$\left( 2002\right) $

\bibitem{Graboske73}  H.C.Graboske, H.E.DeWitt, {\bf 181}$\left( 1973\right)
457$

\bibitem{Ichimaru}  S.Ichimaru, Rev.Mod.Phys. Vol.65, {\bf No.2 }$\left(
1993\right) 255$

\bibitem{adelberger}  E.G.Adelberger et al, Rev.Mod.Phys. {\bf 70}, 1265$%
\left( 1998\right) $

\bibitem{tycholiolios}  T.E.Liolios, {\it nucl-th/0302021}, submitted

\bibitem{Timmes}  F. Timmes, D.Arnett, ApJS, {\bf 125}$\left( 1999\right) $%
277

\bibitem{Iglesias96}  C.Iglesias, F.J.Rogers, ApJ, ${\bf 464}\left(
1996\right) 943$

\bibitem{Kurucz91}  R.L.Kurucz, in ''Stellar Atmospheres:Beyond Classical
Models'', NATO\ ASI Series C, Vol 341 $1991$

\bibitem{AG89}  E.Anders, N.Grevesse, Geochim.Cosmochim. Acta ${\bf 53}%
\left( 1989\right) 197$

\bibitem{Grevesse98}  N.Grevesse, A.J.Sauval, Space Science Reviews, ${\bf 85%
}\left( 1998\right) 161$

\bibitem{Lodders}  K.Lodders, ApJ ${\bf 591}\left( 2003\right) 1220$

\bibitem{Arnettbook}  D.Arnett in ''Supernovae and Nucleosynthesis'',
Prineton University Press $\left( 1996\right) $, ISBN 0-691-01148-6

\bibitem{Maeder}  A.Maeder and G.Meynet, Astron.Astrophys. ${\bf 210}%
(1989)155$

\bibitem{Ribas}  I.Ribas, C.Jordi, J.Torra, A.Gimenez, MNRAS$\,{\bf 313}%
\left( 2000\right) 99$
\end{references}
\end{document}